\documentclass{article}
\usepackage{amsmath, amssymb, amsfonts}
\title{ Gravity and the Superposition Principle } 
\author{Hristu Culetu \footnote{e-mail : hculetu@yahoo.com} \\Ovidius University, Constanta, Romania}

\begin{document}
\numberwithin{equation}{section}
\pagenumbering{arabic}
\maketitle
\newcommand{\fv}{\boldsymbol{f}}
\newcommand{\tv}{\boldsymbol{t}}
\newcommand{\gv}{\boldsymbol{g}}
\newcommand{\OV}{\boldsymbol{O}}
\newcommand{\wv}{\boldsymbol{w}}
\newcommand{\WV}{\boldsymbol{W}}
\newcommand{\NV}{\boldsymbol{N}}
\newcommand{\hv}{\boldsymbol{h}}
\newcommand{\yv}{\boldsymbol{y}}
\newcommand{\RE}{\textrm{Re}}
\newcommand{\IM}{\textrm{Im}}
\newcommand{\rot}{\textrm{rot}}
\newcommand{\dv}{\boldsymbol{d}}
\newcommand{\grad}{\textrm{grad}}
\newcommand{\Tr}{\textrm{Tr}}
\newcommand{\ua}{\uparrow}
\newcommand{\da}{\downarrow}
\newcommand{\ct}{\textrm{const}}
\newcommand{\xv}{\boldsymbol{x}}
\newcommand{\mv}{\boldsymbol{m}}
\newcommand{\rv}{\boldsymbol{r}}
\newcommand{\kv}{\boldsymbol{k}}
\newcommand{\VE}{\boldsymbol{V}}
\newcommand{\sv}{\boldsymbol{s}}
\newcommand{\RV}{\boldsymbol{R}}
\newcommand{\pv}{\boldsymbol{p}}
\newcommand{\PV}{\boldsymbol{P}}
\newcommand{\EV}{\boldsymbol{E}}
\newcommand{\DV}{\boldsymbol{D}}
\newcommand{\BV}{\boldsymbol{B}}
\newcommand{\HV}{\boldsymbol{H}}
\newcommand{\MV}{\boldsymbol{M}}
\newcommand{\be}{\begin{equation}}
\newcommand{\ee}{\end{equation}}
\newcommand{\ba}{\begin{eqnarray}}
\newcommand{\ea}{\end{eqnarray}}
\newcommand{\bq}{\begin{eqnarray*}}
\newcommand{\eq}{\end{eqnarray*}}
\newcommand{\pa}{\partial}
\newcommand{\f}{\frac}
\newcommand{\FV}{\boldsymbol{F}}
\newcommand{\ve}{\boldsymbol{v}}
\newcommand{\AV}{\boldsymbol{A}}
\newcommand{\jv}{\boldsymbol{j}}
\newcommand{\LV}{\boldsymbol{L}}
\newcommand{\SV}{\boldsymbol{S}}
\newcommand{\av}{\boldsymbol{a}}
\newcommand{\qv}{\boldsymbol{q}}
\newcommand{\QV}{\boldsymbol{Q}}
\newcommand{\ev}{\boldsymbol{e}}
\newcommand{\uv}{\boldsymbol{u}}
\newcommand{\KV}{\boldsymbol{K}}
\newcommand{\ro}{\boldsymbol{\rho}}
\newcommand{\si}{\boldsymbol{\sigma}}
\newcommand{\thv}{\boldsymbol{\theta}}
\newcommand{\bv}{\boldsymbol{b}}
\newcommand{\JV}{\boldsymbol{J}}
\newcommand{\nv}{\boldsymbol{n}}
\newcommand{\lv}{\boldsymbol{l}}
\newcommand{\om}{\boldsymbol{\omega}}
\newcommand{\Om}{\boldsymbol{\Omega}}
\newcommand{\Piv}{\boldsymbol{\Pi}}
\newcommand{\UV}{\boldsymbol{U}}
\newcommand{\iv}{\boldsymbol{i}}
\newcommand{\nuv}{\boldsymbol{\nu}}
\newcommand{\muv}{\boldsymbol{\mu}}
\newcommand{\lm}{\boldsymbol{\lambda}}
\newcommand{\Lm}{\boldsymbol{\Lambda}}
\newcommand{\opsi}{\overline{\psi}}
\renewcommand{\tan}{\textrm{tg}}
\renewcommand{\cot}{\textrm{ctg}}
\renewcommand{\sinh}{\textrm{sh}}
\renewcommand{\cosh}{\textrm{ch}}
\renewcommand{\tanh}{\textrm{th}}
\renewcommand{\coth}{\textrm{cth}}

\begin{abstract}
The relation between gravity and quantum mechanics is investigated in this work. The link is given by the wave packet expansion process, rooted from the Uncertainty Principle. The basic idea is to express the de Broglie wavelength used by Schrodinger for a massive particle in terms of the associated Compton wavelength which is replaced by the Michell-Laplace  radius $Gm/c^{2}$ of the spherical object of mass $m\geq m_{P}$, where $m_{P}$ is the Planck mass.

The wave packet spreading is studying in spherical coordinates, having the width $\sigma(t)$, expressed in terms of $G$ and $c$ instead of $\hbar$. Therefore, for masses larger than the Planck mass, a faster dispersion rate of $\sigma(t)$ is obtained, compared to the standard case. The dispersion of the wave packet is observed only by a free falling observer and the process breaks down once the observer hits the surface of the object. Different freely falling observers notice different rates of expansion of the wave packet and the source of gravity is in a quantum superposition. We further confront the Mita formula for the mean energy of the wave packet with the de Broglie-Bohm quantum potential energy when the Schrodinger equation is expressed in the Madelung form. 
 
 \end{abstract}
 
\section{Introduction}
The  search for a consistent and testable theory of quantum gravity is among the most important open problems of fundamental physics. Gravitation is the oldest of the known interactions and, moreover, the most mysterious one \cite{CK}. The fields in the Standard model all carry energies and so generate a gravitational field. Being quantum fields, they cannot be inserted directly into the classical Einstein field equations. Only a full unification of gravity with quantum theory can describe the interaction of fields at the fundamental
level. 
 
  According to Kiefer \cite{CK}, quantum gravity means any theory where the quantum Superposition Principle (SP) is applied to the gravitational field. That is so because the SP (which has been confirmed by a huge number of experiments) is at the heart of quantum theory. Usually, the gravitational field of an object is described by a spatial superposition at different locations. One has to comprise gravity into the quantum framework, since the quantum fields of the non-gravitational interactions represent sources for gravitational field. 
 However, there are no logical arguments that would force us to quantize gravity and hybrid theories can indeed be constructed \cite{AKR}.

  For Marletto and Vedral \cite{MV}, quantum effects in the gravitational field are very small. They adopted a quantum information approach to testing quantum gravity using two masses, each in a superposition of two locations. They proved that any system (for example, a field), mediating entanglement between two quantum systems ought to be quantum. Foo et al. \cite{FAZM} investigated the quantum superposition of different spacetimes not related by a global coordinate transformation - the so-called ''spacetime superpositions''. Their purpose was to study the effects induced on quantum matter residing within such spacetimes. Aspermeyer et al. \cite{ABGM} focused on ''gravitational quantum physics'', an emerging field of research in which phenomena require both quantum theory and gravity, for their explanation.
	
	Konishi \cite{KK} considers that the centre of mass (CM) of a macroscopic object is treated as if it were a pure state described by a wave function, neglecting certain microscopic quantum processes. These microscopic physical processes represent inessential backgrounds and small corrections. He also stressed the role played by object's temperature in cancelling the coherent superposition of macroscopically distinct states.
	
	  Calmet and Hsu \cite{CH} stated that the black hole (BH) information is encoded in entangled macroscopic superposition states of the Hawking radiation, which retains a memory of the original matter configuration that collapsed to a black hole. They further consider a burning lump of coal instead of an evaporating black hole and conclude that the
initial coal state evolves into a macroscopic superposition of radiation states.

The purpose of the present work is to investigate a possible connection between Newtonian Gravity (NG) and nonrelativistic Quantum Mechanics (QM) by means of the wave packet expansion phenomenon. The idea is to replace the Compton wavelength from the Schrodinger equation with a characteristic radius $r_{g} = Gm/c^{2}$ of the spherical object, when its mass is bigger than the Planck mass $m_{P}$. The Newton constant  $G$ and the velocity of light $c$ are rooted from NG and the Maxwell Equations, respectively. The above length has been introduced long time ago by J. Michell and P. Laplace \cite{MOW}. They related it to the radius of a spherically symmetric body at which the escape velocity is equal to $c$.
	
	The above recipe is equivalent with the introduction in the Schrodinger equation of the fundamental constants $G$ and $c$ , instead of the Planck constant $\hbar$. Therefore, the width of the wave packet corresponding to the spherical source of mass $m$ will depend on the Newtonian acceleration $Gm/\sigma_{0}^{2}$, where $\sigma_{0}$ is the value of the wave packet width at $t = 0$. On the grounds of the above slight modification of the Schrodinger equation we then investigate its consequences related to the quantum superposition principle. Using the well-known property of the NG that radial pulsations of the spherical source preserve the static feature of the gravitational field outside it, we conjecture that the macroscopic source of gravity is in superposition of scales, the rate of spreading of the object being coordinate dependent. The spreading is measured only by free-falling observers, defined as observers who feel no other force than gravity (in the Newtonian sense). The free-falling motion is viewed by a static observer on the surface.
	
	We are working in the framework of Newtonian Gravity, using velocity of light from electromagnetism, and so the Schrodinger equation will be written in flat space.
					
	\section{Stationary Schrodinger equation for $m\geq m_{P}$}
	It is instructive to see how Schrodinger guessed his equation. He used the Planck energy formula for a photon $E = \hbar \omega$, where $E$ is the photon energy and $\omega$ its frequency, and the de Broglie expression relating the momentum $p$ of a particle with its associate wavelike quantity, its wavelength $\lambda$: $p = h/\lambda$. Schrodinger started with the wave equation for a massless particle
	   \begin{equation}
		\nabla^{2}f(\textbf{x},t) - \frac{1}{v^{2}}\frac{\partial^{2}f(\textbf{x},t)}{\partial t^{2}} = 0
 \label{2.1}
 \end{equation}
($v$ is the wave velocity) and applied it to the de Broglie waves. With $f(\textbf{x},t) = e^{-i\frac{2\pi v}{\lambda}t} \Psi (\textbf{x})$, one obtains for $\Psi (\textbf{x})$
	   \begin{equation}
		\nabla^{2}\Psi(\textbf{x}) + \frac{4\pi^{2}}{\lambda^{2}}\Psi(\textbf{x}) = 0.
 \label{2.2}
 \end{equation}
For to describe the motion of a massive particle (say, an electron), the above $\lambda$ should be replaced by the de Broglie wavelength $\lambda = 2\pi \hbar /p$. For a free particle we have $E = p^{2}/2m$ such that (2.2) becomes
	   \begin{equation}
		\nabla^{2}\Psi(\textbf{x}) + \frac{2mE}{\hbar^{2}}\Psi(\textbf{x}) = 0,
 \label{2.3}
 \end{equation}
which is the standard stationary Schrodinger equation for a free particle.

Or aim now is to ''adjust'' the above equation for masses $m\geq m_{P}$. Let us write the expression of the de Broglie (dB) wavelength in the following form
	   \begin{equation}
		\lambda_{dB} = \frac{h}{mv} = \frac{h}{mc} \frac{c}{v} = \frac{2\pi \hbar}{mc} \frac{c}{v},
 \label{2.4}
 \end{equation}
where $c$ is the velocity of light in vacuo. One notices that the Compton wavelength $\lambda_{C} = \hbar/mc$ came out in the above equation. For elementary particles in microphysics it is reasonable to use it but that is not so in macrophysics. For instance, $\lambda_{C}$ is completely negligible for the mass of the Moon. Therefore, it is in our opinion more appropriate to replace the Compton wavelength in (2.4) with $r_{g} = Gm/c^{2}$  associated to the macroscopic mass $m$, larger than the Planck mass. In this case $\lambda_{dB}$ from (2.4) acquires the form
	   \begin{equation}
		\lambda_{dB} = \frac{2\pi Gm^{2}}{cp}. 
 \label{2.5}
 \end{equation}
Once (2.5) is introduced in (2.2) we get
	   \begin{equation}
		\nabla^{2}\Psi(\textbf{x}) + \frac{2c^{2}E}{G^{2}m^{3}}\Psi(\textbf{x}) = 0.
 \label{2.6}
 \end{equation}
In other words, for masses bigger than the Planck mass, we replace everywhere $\hbar$ by $Gm^{2}/c$. Two new fundamental constants arose here: the Newton constant $G$ and the velocity of light $c$, instead of $\hbar$. Of course, with $m = m_{P} = 10^{-5}$ grams, one obtains $\hbar = Gm^{2}/c$.

\section{Wave packet expansion}
Let us see now how the expression of the wave packet spreading phenomenon for a free spherical object looks like, with the new fundamental constants inserted in the Schrodinger equation. The well known standard radial probability density in spherical coordinates appears as \cite{BLM, GG}
	   \begin{equation}
		\rho(r,t) = |\Psi (r,t)|^{2} = \frac{e^{-\frac{r^{2}}{\sigma_{0}^{2}\left(1 + \frac{\hbar^{2}t^{2}}{m^{2}\sigma_{0}^{4}}\right)}}}{\pi^{3/2}\sigma_{0}^{3}\left(1 + \frac{\hbar^{2}t^{2}}{m^{2}\sigma_{0}^{4}}\right)^{3/2}},
 \label{3.1}
 \end{equation}
where $\sigma_{0}$ is the width of the wave packet at $t = 0$ and
	   \begin{equation}
		\sigma(t) = \sigma_{0}\sqrt{1 + \frac{\hbar^{2}t^{2}}{m^{2}\sigma_{0}^{4}}}.
 \label{3.2}
 \end{equation}
Inserting $Gm^{2}/c$ instead of $\hbar$ in the above equation, one obtains
	   \begin{equation}
		\sigma(t) = \sigma_{0}\sqrt{1 + \frac{1}{c^{2}}\left(\frac{Gm}{\sigma_{0}^{2}}\right)^{2}t^{2}},
 \label{3.3}
 \end{equation}
where $a \equiv Gm/\sigma_{0}^{2}$ is the Newtonian acceleration at the distance $\sigma_{0}$ from the center of mass (CM) of the macroscopic spherical source with $m\geq m_{P}$, namely the origin of the coordinates. 

Let us consider $\sigma_{0} = R_{0}$, the radius of the macroscopic object at $t = 0$. When (3.3) is written in terms of $R(t)$ and the acceleration $a$, we have
	   \begin{equation}
		R(t) = R_{0} \sqrt{1 + \frac{a^{2}t^{2}}{c^{2}}}, 
 \label{3.4}
 \end{equation}
which represents a hyperbolic motion. 

We take now a static observer $O$, at rest w.r.t. an observer on the surface of the object of radius $R_{0}$, at a height $r_{0}>R_{0}$ (say, at the top of a tower, when we are on the Earth), measured from the origin of spherical coordinates. From $r_{0}$ the observer $O$ drops radially a closed box. Another observer $O'$ inside the box is free-falling (defined to the end of Sec.1) and, from his/her point of view the source of gravity behaves as a matter wave and so it is subjected to spreading (as a free electron, for example, when its wave properties start to manifest), according to (3.4). This is in accordance with Konishi \cite{KK} who stated that the CM of a macroscopic object is treated as if it were a pure state described by a wave function, neglecting microscopic quantum processes.  

From the point of view of $O'$ the expansion of the source ends when the surface of radius $R(t)$ (which is approaching $O'$ according to (3.4)) hits the box and so $O'$ becomes a static observer (the collapse of the wave function of the source takes place). This is equivalent with the moment when the electron from the double-slit experiment hits the screen and behaves like a particle (it is no longer free).  

 With respect to a static observer located on the surface of radius $R_{0}$, the equation of motion of $O'$ is given by
	   \begin{equation}
		r(t) = r_{0} + \frac{c^{2}}{a}\left(1 - \sqrt{1 + \frac{a^{2}t^{2}}{c^{2}}}\right),
 \label{3.5}
 \end{equation}
with $r(0) = r_{0}$. Our aim now is to find the time $T$ when $r(T) = R_{0}$, namely when $O'$ hits the surface. Eq. 3.5 yields
	   \begin{equation}
	 b = \frac{c^{2}}{a}\left(\sqrt{1 + \frac{a^{2}T^{2}}{c^{2}}} - 1\right)            
 \label{3.6}
 \end{equation}
where $b \equiv r_{0} - R_{0}$. For velocities $aT<<c$ we get
	   \begin{equation}
		b \approx \frac{c^{2}}{a}\left(1 + \frac{a^{2}T^{2}}{2c^{2}} - 1\right) = \frac{aT^{2}}{2},
 \label{3.7}
 \end{equation}
which is the well-known Newtonian result. This would represent a confirmation of the validity of the procedure used above. 

It is worth finding the difference between the exact expression (3.6) of $b$ and its Newtonian approximate expression (3.7) which is valid for low velocities. One obtains
	   \begin{equation}
	 \Delta b = \frac{c^{2}}{a}\left(\sqrt{1 + \frac{a^{2}T^{2}}{c^{2}}} - 1\right) - \frac{aT^{2}}{2} \approx \frac{c^{2}}{a}\left(\frac{a^{2}T^{2}}{2c^{2}} - \frac{a^{4}T^{4}}{8c^{4}}\right) - \frac{aT^{2}}{2} = -\frac{a^{3}T^{4}}{8c^{2}} .     
 \label{3.8}
 \end{equation}
If we take the case of the Earth, with $a = 980 cm/s^{2},~T = 10 s,~c = 3\cdot 10^{10}cm/s$, one obtains $\Delta b  \approx -1.3\cdot 10^{-9}cm$. That difference $\Delta b$ could be checked experimentally using the Einstein-Elevator \cite{LFWOE} from the Leibniz Universitat Hannover. One notes that Einstein proposed that experiment years before the General Relativity was born.

Gravity seems to be nothing but a wave packet expansion phenomenon, at least when the tidal forces are negligible \cite{HC2}. Note that the rate of spreading $d\sigma(t)/dt$ is increasing with time. In other words, longer time means increased rate. Therefore, from different location a free falling observer sees different sources with, of course, the same mass $m$. As we know from Newtonian gravity, radial pulsation of the source does not change the static state  outside. We have here a quantum superposition of the source, not in position but in scale (or expansion rate). As Zwirn has noticed \cite{HZ}, each observer builds his/her own reality to which no other observer has any access.

Related to the subject of the paper, the time elapsed until the width $\sigma (t)$ becomes twice its initial value and the mean value of the energy of the wave packet are investigated in Appendix A. The properties of the de Broglie-Bohm quantum potential $Q$ are studied in Appendix $B$ and the Generalized Uncertainty Principle (GUP) is examined in Appendix $C$.

\section{Concluding remarks}
The interaction of fields at the fundamental level can be described only after a full unification of gravity with quantum mechanics. Usually, the quantum effects of the gravitational field of a body are studied by means of a spatial superposition at different locations. We report in this paper a different type of superposition: it is felt only by free falling observers who measure different rates of expansion of the gravitational source, viewed as a wave packet spreading process (we are dealing with a superposition ''in scale'').
 
For $m\geq m_{P}$, we replaced $\hbar$ in the Schrodinger equation with $Gm^{2}/c$, a change coming from the emergence of $r_{g}$ in lieu of the Compton wavelength $\lambda_{C}$. The wave packet spreading of the source is analysed in spherical coordinates and the Newtonian acceleration $Gm/\sigma_{0}^{2}$ is identified in the expression of the time evolution of the width of the wave packet. 

We confronted the Mita formula for the energy of the wave packet in Appendices $A$ and $B$ with the de Broglie - Bohm quantum potential energy from the Madelung form of the Schrodinger equation. A direct connection between the Generalized Uncertainty Principle (GUP) and the role played by gravity is established in Appendix $C$, showing that the 2nd term from the GUP might be interpreted as having a gravitational origin, being important for masses bigger than the Planck mass.

We stress also that the paper uses two constants in lieu of $\hbar$: $G$ from Newtonian Gravity and $c$ from Maxwell Equations.\\

\textbf{Acknowledgements}
 
I am grateful to one of the anonymous referees for useful suggestions and comments which considerably improved the quality of the manuscript.

\section{Appendix A}
Let us compare now the time intervals after which the width $\sigma(t)$ is twice the initial value $\sigma_{0}$. Inserting $\sigma(t) = 2\sigma_{0}$ in (3.2), we get 
	   \begin{equation}
		t_{q} = \frac{\sqrt{3}~m\sigma_{0}^{2}}{\hbar}.
 \label{5.1}
 \end{equation}
 But from (3.3) one obtains
	   \begin{equation}
		t_{g} = \frac{\sqrt{3}~c\sigma_{0}^{2}}{Gm}.
 \label{5.2}
 \end{equation}
With a mass, say, $m = 10^{6}$ grams and $\sigma_{0} = 10^{2}$ cm, we have $t_{q} \approx 10^{37}$s (much more than the age of the Universe) and $t_{g} \approx 10^{16}$s, less than the age of the Universe. When $m = m_{P}$ we obtain, of course, $t_{q} = t_{g} \approx 10^{26}$s.

We wish now to comment on the mean value of the energy of the wave packet. It was calculated by Mita \cite{KM}. His expression of the mean energy in one spatial dimension is $<p^{2}/2m> = mv_{0}^{2}/2 + \hbar^{2}/2m\sigma_{0}^{2}$. We have in our situation $v_{0} = 0$ (no linear motion) and, for $m>m_{P}$ 
 	   \begin{equation}
		<\frac{p^{2}}{2m}> = \frac{mc^{2}}{2}\left(\frac{r_{g}}{\sigma_{0}}\right)^{2}
 \label{5.3}
 \end{equation}
Mita \cite{KM} designated its energy expectation value as ''dispersion oscillations'' of the particle, or energy of localization, without giving a precise nature of those oscillations.  
Our expression (5.3) for the mean energy could be interpreted as ''spreading energy'' which becomes half of the rest energy when $R_{0} = \sigma_{0} = r_{g}$. 

\section{Appendix B}
It is instructive to find out another type of energy, the so called de Broglie-Bohm quantum potential energy $Q$ \cite{PH, LPT}. Holland \cite{PH} suggested that $Q$ may be regarded as the kinetic energy of additional ''concealed'' degrees of freedom. It is obtained from the Madelung form of the Schrodinger equation, when the wave function is written as $\Psi = \sqrt{\rho}~exp(iS/\hbar)$, where $\rho$ is the amplitude squared and $S$ is the phase of the wave function. One equation is the continuity equation and the other is equivalent with the Hamilton-Jacobi equation, but with an extra term given by
 	   \begin{equation}
		Q = -\frac{\hbar^{2}}{2m}\frac{\nabla^{2}\sqrt{\rho}}{\sqrt{\rho}}.
 \label{6.1}
 \end{equation}
It is the only term from the two equations containing $\hbar$. According to Esposito \cite{SE} (see also \cite{PH}), $Q$ is a kinetical energy for an internal motion of the object, the external motion being interpreted as the motion of the CM. 
Using $\rho$ from (3.1) it can be shown that, when $m>m_{P}$, $Q$ is given by
 	   \begin{equation}
		Q(r,t) = \frac{mc^{2}}{2}\left(\frac{r_{g}}{\sigma(t)}\right)^{2}\left[3 - \frac{r^{2}}{\sigma^{2}(t)}\right].
 \label{6.2}
 \end{equation}
When $Q(r,t)$ is calculated in terms of $\hbar$, one obtains the same expression as that provided by Rahmani and Golshani \cite{RG}. The number ''3'' within the square parantheses comes from the three spatial dimensions (spherical coordinates). With one spatial dimension $x$ we have ''1'' instead of ''3'' and $x$ instead of $r$. It is clear that the Mita energy is just the quantum potential energy $Q$ at $r = 0,~t = 0$, when $Q$ equals $Q_{max} = mc^{2}r_{g}^{2}/2\sigma_{0}^{2}$ (in one spatial dimension). 

We persuade ourselves of that by computing the expectation value of $Q_{0}$ at the initial time $t = 0$ or for a stationary state when the probability density of a quantum system does not depend on time. We have
 	   \begin{equation}
		<Q_{0}> \equiv <Q(r,0)> = \int{\Psi^{*}(r,0)Q_{0}\Psi (r,0)dV}, 
 \label{6.3}
\end{equation}
where $dV$ is the volume element in spherical coordinates. From (6.3) one obtains
 	   \begin{equation}
		<Q_{0}> = \frac{1}{\pi \sqrt{\pi}\sigma_{0}^{3}}~\frac{mc^{2}}{2}~\frac{r_{g}^{2}}{\sigma_{0}^{2}} \int_{0}^{\infty}\left(3 - \frac{r^{2}}{\sigma_{0}^{2}}\right)e^{-\frac{2r^{2}}{\sigma_{0}^{2}}}~ 4\pi r^{2}dr
 \label{6.4}
\end{equation}
With the help of the well-known relations
 	   \begin{equation}
	\int_{0}^{\infty}y^{2}e^{-\frac{y^{2}}{b^{2}}} dy = \frac{\sqrt{\pi}}{4}b^{3},~~~~~~\int_{0}^{\infty}y^{4}e^{-\frac{y^{2}}{b^{2}}} dy = \frac{3\sqrt{\pi}}{8}b^{5},
 \label{6.5}
\end{equation}
we get from (6.4)
 	   \begin{equation}
		<Q_{0}> = \frac{9}{32\sqrt{2}} ~\frac{r_{g}^{2}}{\sigma_{0}^{2}}~ mc^{2},~~~~m\geq m_{P},
 \label{6.6}
\end{equation}
which, for $\sigma_{0} = r_{g}$ it will be of the order of $ mc^{2}$. For $m < m_{P}, <Q_{0}>$ acquires the form
 	   \begin{equation}
		<Q_{0}> = \frac{9}{16\sqrt{2}} \frac{\hbar^{2}}{2m\sigma_{0}^{2}} ,
 \label{6.7}
\end{equation}
which has the same form as the Mita ''localization energy'' $E_{L}$. In our situation, $<Q_{0}>$ is not a kinetic energy but a potential one, that will become kinetic from the point of view of a free falling observer. Let us exhibit an estimation of the mean value of $Q_{0}$ from (6.6). Take for $m$ the mass of the Earth, $m\approx 6\cdot 10^{27}$ grams, $r_{g} = 2.5 ~cm$, and $\sigma_{0} = 6.37 \cdot 10^{8}$ cm. With these values, one obtains $<Q_{0}> \approx 10^{31}$ ergs, a reasonable value. If one uses the formula (6.7) for the same mass, we get a very tiny value, completely negligible.

From $Q(r,t)$ the expression of the quantum force appears as
 	   \begin{equation}
		\textbf{F}_{Q} = -\nabla Q = -\frac{\partial Q(r,t)}{\partial r}\textbf{e}_{r} = \frac{mc^{2}r_{g}^{2}}{\sigma^{4}(t)}\textbf{r}.
 \label{6.8}
 \end{equation}
Note that the quantum force (which is repulsive, being positive) is proportional to $\textbf{r}$, as the expansion force in the case of the de Sitter Universe in static coordinates.

\section{Appendix C}
As far as the Uncertainty Principle is concerned, it is worth noting that the appearance of the Planck constant on the r.h.s. of the Heisenberg uncertainty relation originates from the commutator of two operators, and there from experiments in Microphysics. Therefore, macroscopically we may replace $\hbar$ by its macroscopic counterpart, namely $Gm^{2}/c$. It leads us to
	   \begin{equation}
		\Delta r \Delta p = \frac{\sigma_{0}}{\sqrt{2}}\sqrt{1 + \frac{a^{2}t^{2}}{c^{2}}}\frac{\hbar}{\sqrt{2}\sigma_{0}}\geq \frac{\hbar}{2} ~~\rightarrow ~~ \frac{Gm^{2}}{2c} = \frac{1}{2}r_{g} mc,
 \label{7.1}
 \end{equation}
with $r_{g} \approx (\Delta r)_{min}$. 
We notice that the momentum $mc$ in (7.1) plays the role of the constant $\Delta p \approx \hbar/\sigma_{0}$ and we get the maximal value $\Delta p = mc$  when $\sigma_{0}$ equals the Compton wavelength of the particle or its radius $r_{g}$ when $m\geq m_{P}$.

If we keep track of the Generalized Uncertainty Principle ($GUP$) \cite{SC} (see also \cite{CR, PB}) 
	   \begin{equation}
		\Delta x \geq \frac{\hbar}{\Delta p} + l_{P}^{2}\frac{\Delta p}{\hbar},
 \label{7.2}
 \end{equation}
where $l_{P} = 10^{-33}$cm is the Planck length, one observes that the 2nd term of the r.h.s. of (7.2) does not depend on $\hbar$, so it appears as
	   \begin{equation}
		\Delta x \geq \frac{\hbar}{\Delta p} + \frac{G}{c^{3}}{\Delta p},
 \label{7.3}
 \end{equation}
If we look for $(\Delta x)_{min}$, it is obtained when $\Delta p = mc$, i.e., the value from (7.1). Taking into consideration that, in our view, for $m\geq m_{P} = 10^{-5}$grams, $\hbar \rightarrow Gm^{2}/c$, Eq.(7.3) yields 
	   \begin{equation}
		(\Delta x)_{min}= \frac{2Gm}{c^{2}},
 \label{7.4}
 \end{equation}
due to the contribution from both terms.
Eq. (7.3) may be also written as
\begin{equation}
		\Delta x \Delta p \geq \hbar \left[1 + \left(\frac{l_{P} \Delta p}{\hbar}\right)^{2}\right].
 \label{7.5}
 \end{equation}
We distinguish here two situations:

i) if $\Delta p << \hbar /l_{P} = m_{P}c$, ~~~$\Delta x \Delta p \geq \hbar$ .

ii) if $\Delta p >> \hbar /l_{P} = m_{P}c$, ~~~$\Delta x \Delta p \geq (G/c^{3})(\Delta p)^{2}$, or ~~~$\Delta x \geq (G/c^{3})\Delta p$,

that is always valid if $\Delta x \geq (G/c^{3})(\Delta p)_{max} = Gm/c^{2}$ (the factor of 2 is missing because we started from Carlip's \cite{SC} equation (7.2)).
As we anticipated before, the Planck mass (or the Planck momentum) decides which term is more important in the r.h.s. of (7.3). The gravitational term (depending on $G$) dominates if the mass of the object is bigger than the Planck mass.


\begin{thebibliography} {21}

\bibitem{CK}
C. Kiefer, Quantm Gravity, Oxford University Press (2012).
\bibitem {AKR}
M. Albers et al., Phys. Rev. D78, 064051 (2008) (arXiv: 0802.1978).
\bibitem{MV}
C. Marletto and V. Vedral, Phys. Rev. Lett. 119 (2017) 240402.
\bibitem{FAZM}
J. Foo et al., Phys. Rev. D107 (2023) 4, 045014 (arXiv: 2208.12083).
\bibitem{ABGM}
M. Aspelmeyer et al., New J. Phys. 19 (2017) 050401.
\bibitem{KK}
K. Konishi, Int. J. Mod.Phys. A38 (2023) 14, 2350080 (arXiv: 2209.07318). 
\bibitem{CH}
X. Calmet and S. Hsu, Europhys. Lett. 139 (2022) 4, 49001 (arXiv: 2207.08671).
\bibitem{MOW}
C. Montgomery,,  W.Orchiston and I. Whittingham, J. of Astron. Hist. and Heritage 12(2) 90 (2009).
\bibitem {BLM}
L. Buoninfante et al., Eur. Phys. J. C78 (2018) 1, 73 (arXiv: 1709.09263).
\bibitem{GG}
D. Giulini and A. Grossardt, Class. Quantum Grav. 28 (2011) 195026 (arXiv: 1105.1921).
\bibitem{LFWOE}
C. Lotz et al., Grav. Space Research 5(2), 11 (2017).
\bibitem{HC2}
H. Culetu, Short Communication presented at the Meeting ''Quantum simulations of gravitational problems on condensed matter analog models'', Trento, Italy, June 19 - 23, 2023.
\bibitem{HZ}
H. Zwirn, Quantum Rep. 5 (2023) 1, 267 (arXiv: 2301.07532).
\bibitem{KM}
K. Mita, Am. J. Phys. 71, 894 (2003).
\bibitem{PH}
P. Holland, Found. Phys. 45, 134 (2015) (arXiv: 1410.0165).
\bibitem{LPT}
H. S. Lima, M. Paixao and C. Tsallis, arXiv: 2308.07178.
\bibitem{SE}
S. Esposito, Found. Phys. Lett. 12, 165 (1999) (arXiv: 9902019).
\bibitem{RG}
F. Rahmani and M. Golshani, Int. J. Mod. Phys. A36 (2021) 26, 2150181 (arXiv: 2012.04606).
\bibitem{SC}
S. Carlip, Rept. Prog. Phys. 64, 885 (2001) (arXiv: gr-qc/0108040).
\bibitem{CR}
C. Rovelli, 9th Marcel Grossmann Meeting MG9, pp. 742 (2000) (arXiv: gr-qc/0006061).
\bibitem{PB}
P. Bosso et al., Class. Qantum Grav. 40 (2023) 19, 195014 (arXiv: 2305.16193).



\end{thebibliography}
\end{document}